\documentclass[epj,nopacs]{svjour}
\usepackage{graphicx}
\usepackage{bm}
\usepackage{mathtools}
\usepackage{amssymb}
\usepackage{color}
\usepackage[colorlinks=true, linkcolor=blue, citecolor=red, urlcolor=blue]{hyperref}
\usepackage{cite}

\DeclarePairedDelimiter\parens{(}{)}
\usepackage[UKenglish]{babel}
\usepackage{microtype}
\providecommand{\red}[1]{\textcolor[rgb]{0,0,0}{#1}}
\newcommand{\g}[1]{\mbox{\boldmath $#1$}}

\begin{document}
\title{Relativistic Vlasov-Maxwell modelling using finite volumes and
  adaptive mesh refinement}
\titlerunning{Relativistic Vlasov-Maxwell modelling with AMR}
\author{B.\ Svedung Wettervik \and T.\ C.\ DuBois \and E.\ Siminos \and T.\ F\"ul\"op}

\institute{Department of Physics, Chalmers University of Technology, Gothenburg, Sweden}
\date{Received: \today / Revised version: date}
%
\abstract{
The dynamics of collisionless plasmas can be modelled by the
Vlasov-Maxwell system of equations.  An Eulerian approach is
  needed to accurately describe processes that are governed by high
  energy tails in the distribution function, but is of limited
  efficiency for high dimensional problems. The use of an adaptive
  mesh can reduce the scaling of the computational cost with the
  dimension of the problem.  Here, we present a relativistic
Eulerian Vlasov-Maxwell solver with block-structured adaptive mesh
refinement in one spatial and one momentum dimension.  The
discretization of the Vlasov equation is based on a high-order
finite volume method. A flux corrected transport algorithm is
  applied to limit spurious oscillations and ensure the physical
  character of the distribution function. We demonstrate a
  speed-up by a factor of $7$x in a typical scenario involving laser pulse interaction 
      with an underdense plasma due to the use of an adaptive mesh.
} 
\maketitle

\section{Introduction}
The Vlasov-Maxwell system of equations describes the dynamics of a
collisionless plasma whose component species interact through
self-consistent electromagnetic fields.  It is of critical
importance to the fundamental understanding of non-equilibrium
processes in collisionless plasmas, as well as many practical
applications, for example laser plasma acceleration
\cite{malka,Daido},
inertial confinement fusion \cite{brunner2004,benisti2010,winjum2013},
high harmonic generation \cite{HHG} and
shocks in astrophysical plasma \cite{astro}.

Numerical approaches to solve this system are primarily divided into
Particle-In-Cell (PIC) methods, which approximate the plasma by a
finite number of macro-particles, and methods that discretize the
distribution function on a grid: so-called Eulerian methods \cite{shoucri}.  As PIC
methods do not require a grid in momentum space, they are efficient at
handling the large range of scales associated with relativistic
laser-plasma interaction. They are therefore suitable for
  modelling high dimensional problems \cite{Besse2008}. However, the
approximation of the distribution function by a finite number of
particles introduces statistical noise, making it difficult to
resolve fine velocity space structures and high energy tails within
the distribution function.

The ability to resolve fine structures related to low density tails in
the distribution function is of critical importance to topics in laser
plasma acceleration, e.g., modelling of collisionless shock
acceleration (CSA).  In laser plasma induced CSA, a small fraction of
the ion population is reflected by an electrostatic potential barrier
set up by laser plasma interaction. This low density tail of the
distribution function has been suggested to play an important role
for the dynamics \cite{Macchi}, indicating that high resolution is
needed.

Eulerian methods, which discretize the distribution function on a grid
have very low levels of numerical noise. They are therefore
appropriate for the detailed study of processes where a small number
of high energy particles play a significant role.
The most widely used method for solution of the Vlasov equation is
time-splitting, first suggested by Knorr and Cheng
\cite{Cheng1976}.
The method involves splitting the Vlasov equation
into lower dimensional advection equations that are alternately
advanced.
Second order accurate time splitting methods have been used
to solve the Vlasov-Poisson and Vlasov-Maxwell system of equations in
Refs.~\cite{Cheng1976,Filbet2001,Arber2002,Ghizzo2003,Sircombe2009,Grassi2016,veritas2016}.
In these references the distribution function is represented on a fixed uniform grid, which leads to large computational costs, especially for multidimensional simulations.

To increase the computational efficiency and be able to treat problems
with a wide range of scales, the Vlasov-Maxwell system can be
represented on an adaptive mesh in combination with higher order
methods.
For this approach, high-order finite volume discretizations can be used to solve the Vlasov-Maxwell equation system, see
for example Ref.~\cite{Banks2010}.
The adaptive
mesh then evolves as the characteristics of the distribution function
develop, which allows higher resolution to be applied to those parts of
phase space that exhibit complicated behaviour.
At the same time, the
distribution function remains well-resolved in regions with a coarse
mesh through the use of high-order numerical schemes.
This means that the use of adaptive meshing limits the computational effort to regions with
small scales but still maintains a high degree of accuracy in the full domain.

Adaptive Mesh Refinement (AMR) is not a new concept, and has been used extensively
to minimise computational overhead in a variety of systems. Some examples focusing
on laser plasma research include: the PIC code Warp~\cite{Vay2002}, which implements AMR for its
 Poisson solver; M. Dorr \emph{et al}.~\cite{Dorr2002}
 investigates speckles and filamentation in inertial confinement fusion
by solving the Poisson-Euler equations coupled to Maxwell
equations; Ref.\ \cite{Arslanbekov2013} which solves the Boltzmann
equations using a hybrid octree--AMR approach.
AMR has also been used in Ref.~\cite{Hittinger2013} to simulate the Vlasov-Poisson
system, i.e.\ the electrostatic and non-relativistic limit of the Vlasov-Maxwell equations.
Finally, N. Besse \emph{et al.} \cite{Besse2008} use a wavelet based adaptive grid to solve the relativistic Vlasov-Maxwell system.

In this paper we present the open source block-structured Eulerian Vlasov-Maxwell solver {\sc veritas}\linebreak
({\bf V}lasov {{\bf
  E}ule{\bf RI}an} {\bf T}ool for {\bf A}cceleration {\bf
    S}tudies) \cite{veritas}. The solver is based on a high-order finite volume
method, implementing the flux corrected transport algorithm to
limit spurious oscillations in the distribution functions,
  in the presence of steep gradients. {\sc veritas} offers the capability to study realistic laser--plasma problems in two dimensions (1D1P), with a complete electrodynamic framework (i.e.\ Vlasov-Maxwell) at relativistic speeds.
  To our knowledge, {\sc veritas} is the first complete, relativistic Vlasov-Maxwell solver using AMR which shows a significant performance increase. This advancement moves continuum solvers towards the category of capable simulation tools alongside their PIC counterparts.

The paper is organized as follows. Section II describes the
Vlasov-Maxwell equations. Section III presents the numerical scheme
for {\sc veritas}, including a description of the mesh structure, information flow, regridding procedure, finite volume discretization of the Vlasov equation and the discretization of Maxwells equations.   Section IV describes benchmarking  with comparison to results  from analytical theory and PIC simulations, and  demonstrates the improved performance of the adaptive mesh approach.  Conclusions
are summarized in Section V.

\section{The Vlasov-Maxwell system}
The Vlasov-Maxwell equation system describes the time evolution of the
electron and ion distribution functions, which interact
self-consistently with the electromagnetic fields in a
  collisionless plasma.  For the case of a plasma with spatial
variation in one direction, the Vlasov-Maxwell equations can be
reduced to a two dimensional 1D1P problem:
\begin{equation}
    \frac{\partial f_s}{\partial t}+\frac{p_x}{m_s\gamma}\frac{\partial
        f_s}{\partial x}+q_s\left[E_x+\frac{1}{\gamma m_s}(\bm{p} \times
        \bm{B})_x\right]\frac{\partial f_s }{\partial p_x}=0,\label{eq:veq}
\end{equation}
where $f_s$ is the distribution function of a species (e.g.\ electrons or ions), $\g E$
  and $\g B$ are the electric and magnetic fields, $x$ is a spatial
coordinate, $p_x$ is a momentum coordinate in the direction of
  $x$, $q$ is the charge, $m$ denotes the rest mass of the
charged particles (electrons or ions) and
$\gamma =\sqrt{\g p^2/m^2c^2+ 1}$ is the relativistic factor.  The
single-particle Hamiltonian
\begin{equation}
   H=mc^2\left[1+\frac{(\bm{\Pi}-q \bm{A})^2}{m^2c^2}\right]^{1/2}+q\phi
\end{equation}
yields conservation relations for the transverse canonical momentum
(orthogonal to the direction of variation of the plasma):
$\bm{\Pi}_\perp=q\bm{A}_\perp+\bm{p}_\perp=0$ \cite{Huot2000,Ghizzo2013}.  The conservation of
$\bm{\Pi}_\perp$ stems from the fact that the perpendicular coordinates $y$ and $z$
do not enter the Hamiltonian.  Here, $c$ is the speed of light and
$\phi$ and $\bm{A}$ are the electrostatic and vector potentials,
respectively.

For a one-dimensional system, Maxwell's equations take the form
\begin{equation*}
\frac{\partial B_x}{\partial x}=0,\quad \frac{\partial B_x}{\partial t}=0,
\end{equation*}
\begin{equation*}
\frac{\partial B_y}{\partial t}=\frac{\partial E_z}{\partial x},\quad \frac{\partial B_z}{\partial t}=-\frac{\partial E_y}{\partial x},
\end{equation*}
\begin{equation*}
\frac{\partial E_x}{\partial x}=\rho/\epsilon_0,\quad \epsilon_0\mu_0\frac{\partial E_y}{\partial t}=-\mu_0J_y-\frac{\partial B_z}{\partial x}
\end{equation*}
and
\begin{equation*}
   \epsilon_0\mu_0\frac{\partial E_z}{\partial t}=-\mu_0J_z+\frac{\partial B_y}{\partial x}.
\end{equation*}
Here, the currents
and charge density are determined by the distribution functions,
according to
\begin{equation}
 \bm{J}_\perp =\sum_s \frac{q_s}{m_s}\int \frac{\bm{p}_{\perp s}}{
  \gamma_s}f_s\,\mathrm{d}p_x\label{eq:current}
\end{equation}
and
\begin{equation}
   \rho=\sum_s q_s \int f_s \,\mathrm{d}p_x,
\end{equation}
where the summation
ranges over all species $s$ in the plasma.  The transverse vector potential
$\bm{A}_\perp$ is obtained by $\bm{E}_\perp=-\partial
\bm{A}_\perp/\partial t$ and is used together with the conservation of
canonical momentum $\bm{\Pi}_\perp$ to calculate the relativistic
factor $\gamma$ and the transverse components of the current.

\section{Numerical scheme}
 The distribution function is represented on a block structured mesh,
 which is adapted during time evolution to ensure that regions of
 phase space with more complex dynamics are associated with a finer
 resolution in the mesh \cite{Berger1984}. We use a finite
 volume scheme which is fourth-order accurate, making it possible to
 use a very coarse representation of the distribution function in
 regions with less complex dynamics, without introducing
 numerical instabilities.

The electromagnetic fields are
defined on a mesh associated with the finest spatial
resolution of the distribution function,
using a fourth-order discretization of Maxwell's equations and a
staggered grid for the electric and magnetic fields. As the fields are one dimensional, the use of a mesh with the finest spatial resolution as opposed to using adaption to adjust the resolution comes at only a moderate cost.

Distribution functions, for one or multiple plasma charge species,
are time advanced together with the electromagnetic fields, yielding
an overall self-consistent solution to the Vlasov-Maxwell
system. In the following, we describe the mesh structure,
the representation of the solution, the
adaption and information flow from one mesh to another, as well as the
discretizations used in {\sc veritas}.

\subsection{Mesh structure}

To resolve the different scales of the distribution function
$f(x,p_x)$, the domain
$D=[x_\text{min},x_\text{max}]\times[p_{x,\text{min}},p_{x,\text{max}}]$
is discretized using a block structured mesh.  The mesh consists of a
number of levels $L_i$ of different resolution, where
$i=0,\dots,n_{\text{max}}$, ranging from coarsest to finest.  Each
level is a union of rectangular patches, i.e.\ $L_i=\cup_{j}R_{i,j}$,
where $R_{i,j}$ denotes a rectangular patch on level $i$.  Each
rectangular patch, which is disjoint from all other rectangular patches on
the same level, consists of a number of rectangular cells of side
lengths $\Delta x_i=\Delta x_0/r^{i}$ and $\Delta p_{x,i}=\Delta
p_{x,0}/r^{i}$, {where $r$ is the refinement ratio.}

The distribution function is described by cell-averaged values
$\tilde f_{k,l}^{m,n}$:
\begin{equation}
\tilde f_{k,l}^{m,n}=\frac{1}{\Delta x_m\Delta
  p_{x,n}}\int_{x_{k-\frac{1}{2}}^m}^{x_{k+\frac{1}{2}}^m}\int_{p_{x,l-\frac{1}{2}}^n}^{p_{x,l+\frac{1}{2}}^n}f(x,p_x)\,\mathrm{d}x\mathrm{d}p_x
\end{equation}
where $m$, $n$ are indices for the level of refinement along the
spatial and momentum dimension, respectively, and $x_{k\pm\frac{1}{2}}^m$,
$p_{x,l\pm\frac{1}{2}}^m$ are the bounding dimensions of the cell.
Throughout this paper we indicate a cell-averaged value with a tilde (\~{}).
Concerning the representation of the distribution function on the block
    structured mesh, the indices $m$ and $n$ always take the same value. The use of cells in the mesh with different $m$ and $n$ would correspond to independently adapting in the $x$ and $p_x$ direction. Although in principle possible, this would complicate the mesh structure and is currently forbidden.
On the other hand, in the calculation of the current and charge densities, the distribution function is interpolated to the finest spatial mesh; which may result in values in cells with different indices for $m$ and $n$.

\begin{figure}[tb]
    \includegraphics[width=\columnwidth]{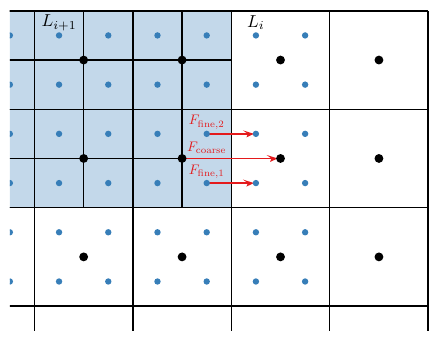}
\caption{\label{fig:cellidx}Coarse rectangular patch on level $L_i$ that
  partially overlaps with finer rectangular patch at level
  $L_{i+1}$. Cell-centers of coarse and fine cells are denoted by black
  and blue dots respectively. The finer rectangular patch has an
  interior region (shaded blue), which is complemented by a
  set of ghost cells (identified by blue dots outside the shaded
  region) used for flux calculations. At a coarse-fine interface, we
  define fluxes $F_{\text{fine,}1}$, $F_{\text{fine,}2}$ and
  $F_{\text{coarse}}$, where $F_{\text{coarse}}$ is inferred from
  $F_{\text{fine,}1}$, $F_{\text{fine,}2}$ in order to ensure particle
  conservation. }
\end{figure}

The coarsest level $L_0$ consists of a single rectangle and finer
levels are { nested} in coarser levels $L_{i+1}\subset L_{i}$.  Figure
\ref{fig:cellidx} shows a rectangular patch on level $L_{i+1}$ which is
partially overlapped by a rectangular patch at level $L_{i}$.  The finer
rectangular patch is shaded blue with a blue dot at the center
of each cell and the coarser cells are indicated by black dots at the
cell centers.  Blue dots outside the blue shaded region represent
ghost cells which are used to time advance the interior cells.  Ghost
cells are interpolated from the cell-centered values in the overlaid
coarser level, as will be described in Section \ref{sec:cfi}, or read from an adjacent rectangle on the same level (i.e.\ interior ghost cells).  Values
of the distribution function on a coarser level which is overlaid by a
finer level, e.g. the black dotted cells inside the blue shaded
region, are defined by the average of the values on the finer level
$\cramped{\tilde f_{k,l}^{i,i} = 1/r^2\sum_{n=0}^{r-1}\sum_{m=0}^{r-1}\tilde
  f^{i+1,i+1}_{rk+n,rl+m}}$.

The division of the mesh into levels and rectangles leads to the
following synchronization procedure at each time step:
\begin{enumerate}
\item From the finest to coarsest level, for each pair of levels
  $L_i$, $L_{i+1}$, use values at the finer level $L_{i+1}$ to
  calculate the values for the overlaid cells in the coarser mesh.
\item For a given rectangle, a ghost cell may either belong to another
  rectangle on the same level or be interpolated from the next coarser
  level. For each level $L_i$, update the value for the ghost cells
  which are interior to the level.
\item From the coarsest to the finest level, for each pair of levels
  $L_i$, $L_{i+1}$, interpolate ghost cells for rectangles in $L_{i+1}$
  which are not interior, using values from $L_i$.
\end{enumerate}

The time stepping procedure itself is outlined in Section \ref{sec:timestep}.

\subsection{Adaptive mesh refinement}

\begin{figure}[tb]
    \includegraphics[width=\columnwidth]{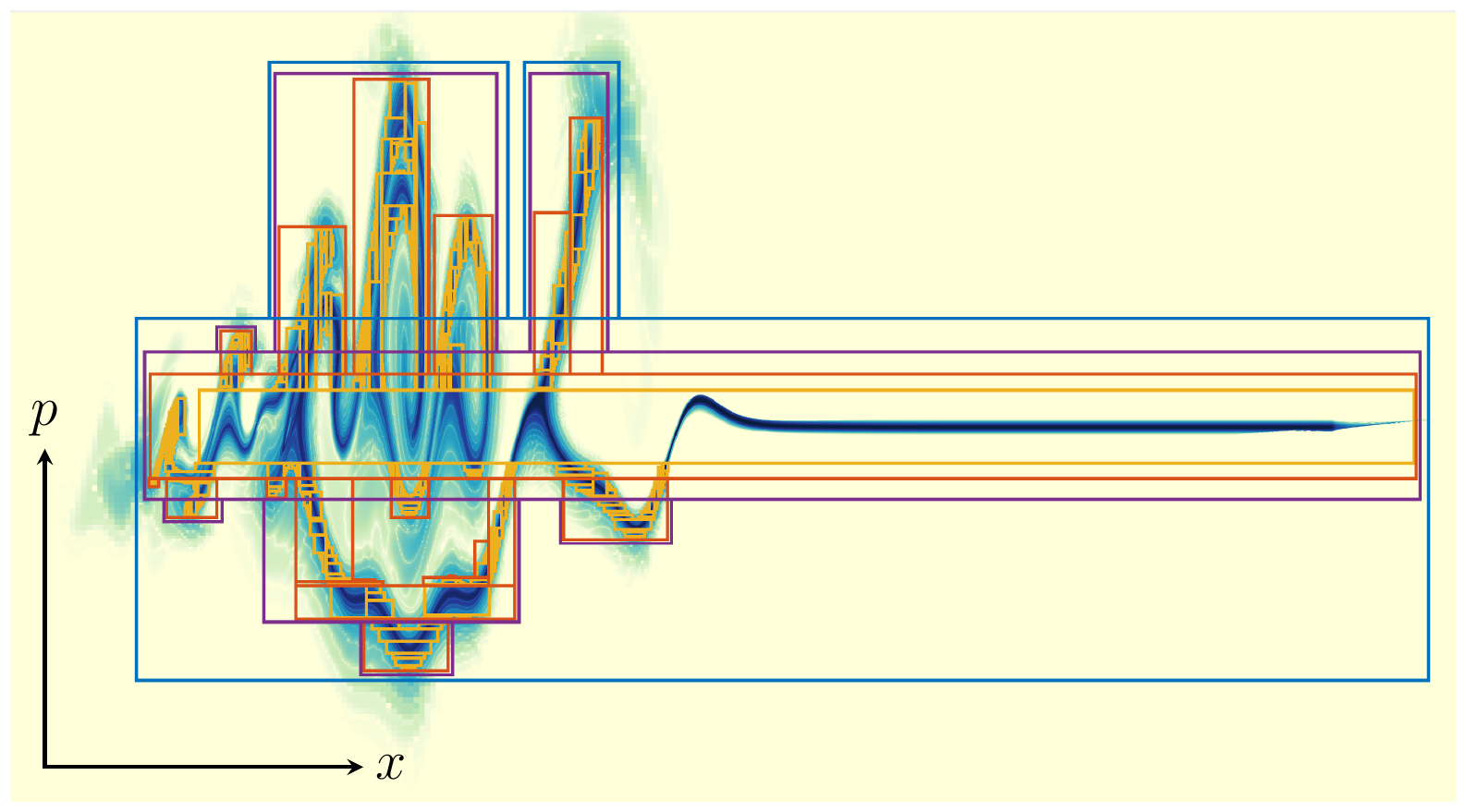}
\caption{\label{fig:adaption}Laser pulse interacting with an electron plasma slab, which is represented using a five level adaptive mesh. The color map shows the 10 largest orders of magnitude for the distribution function on a $\log_{10}$ scale.   The rectangular patches in each level are indicated by colored boxes. The coarsest level is not highlighted via a rectangle, but extends over the entire figure. The levels are nested, with each level (with yellow being the finest)  contained within the next  {coarser} level.}
\end{figure}

Figure \ref{fig:adaption} shows a plasma which has interacted
with a laser pulse. The distribution function is represented on five
levels which are divided into rectangles, with high resolution in
regions with higher densities and more complex dynamics. To track
these regions, the mesh structure is updated in the following way:
\begin{enumerate}
\item For each level $L_i$, mark each {cell} that belongs to a
  rectangle in the level {and} has an error indicator exceeding some
  threshold $\delta_{\rm thres}$, discussed in Section \ref{sec:error}. The marked points $M_i$ are used to generate a
  new level and rectangle structure $L_{i}^{\prime}$ where $i=0,...,
  n^{\prime}_\text{max}$.
\item Let $ n^{\prime}_\text{max}-1$ be the largest integer such that
  $M_{ n^\prime_\text{max}-1}$ is non-empty and create a new level $
  L^{\prime}_{\tilde n_\text{max}}$ such that
  $M_{n_\text{max}^{\prime}-1}\subset  L^{\prime}_{
    n^{\prime}_\text{max}}$.
\item For $i= n_\text{max}^{\prime}-1,...,1$, construct a new level $
  L_{i}^{\prime}$ such that $ L_{i+1}^{\prime} \cup
  M_{i-1}\subset L^\prime_{i}$.
\end{enumerate}

Furthermore, upon adapting the mesh, the representation of the
solution on the old mesh structure must be transformed into a
representation of the solution on the new one.  Information contained
in a cell in the new level $L^{\prime}_i$ may originate from $L_i$ or
a subset of a cell in $L_j$ for some $j<i$.  In the former case, the
cell-averaged value in $L^{\prime}_i$ is taken directly from $L_i$,
whereas in the latter case an interpolation is performed. This operation is
performed for all $L^{\prime}_i$ before interpolating those values
that cannot be copied from $L^{}_i$.  As the coarsest level is the
same in both the old and new mesh, all values in $L^{\prime}_{0}$ are
determined from values in $L^{}_{0}$. Values in cells of
$L^{\prime}_{i}$ that are not a subset of $L_i$ can then be
interpolated from $L^{\prime}_{i-1}$ for $i=1$, ...,
$n^{\prime}_\text{max}$.

The covering of $ M_i^\prime\equiv L_{i+1}^{\prime} \cup M_{i-1}$ with rectangles
closely follows the AMR implementation pioneered by Berger \textit{et
  al.}~\cite{Berger1984, Berger1989}, with specific extensions
outlined in their follow-up papers~\cite{Berger1991, Bell1994}.
To keep dependencies to a minimum, and integration tight \& efficient, no external libraries are called here and the AMR infrastructure outlined below has been implemented directly into {\sc veritas}.
For each level, from
the finest to coarsest, we identify a minimum bounding box
$R_{i,\mathrm{bb}}$,
such that all marked cells on level $i$ are extant within the
rectangle's boundaries.  $R_{i,\mathrm{bb}}$ is then split into
smaller rectangles $R_{i,j}$ through a recursive process until a minimum
efficiency, defined as the proportion of marked cells in each rectangle, is reached:
\begin{equation}
   \frac{N[ M_i^\prime \cap R_{i,j} ]}{N[R_{i,j}]} \geq \varepsilon_{\mathrm{min}}.
\end{equation}
Here, $N[\cdot]$ denotes the number of cells within a given set,
and
$\varepsilon_{\mathrm{min}}$ is a user{-}set parameter for the minimum
efficiency.  {The specific value of the parameter} will affect
the overall runtime of the program.
{ A low value for  $\varepsilon_{\mathrm{min}}$ leads to a larger mesh with more degrees of freedom, with a {corresponding} increase in computational work. On the other hand, a high value leads to a more complicated mesh structure, with larger overheads,  which in addition must be updated more frequently to accurately  represent the solution. Values for  $\varepsilon_{\mathrm{min}}$ between $60$ \% and $80$ \%, are found to give reasonable performance, although optimal values are problem dependent. }

To split a rectangle, we introduce signatures $\Sigma_{x,k}$ and
$\Sigma_{p_x,l}$, which are  functions of the discrete $x$ and
$p_x$ coordinates, respectively. For a given rectangle, $\Sigma_{x,k}$
is defined as the number of cells in the intersection of the rectangle
and marked cells  $M_i^\prime$ that have the spatial index $k$.  $\Sigma_{p_x,l}$ is
defined similarly, but with the spatial and momentum coordinates
exchanged.
A rectangle can be split into two smaller rectangles at an index,
either $k$ or $l$, at which the corresponding signature is zero.  If
neither signatures contain a zero, the signature's derivatives
$\Delta_x$ and $\Delta_{p_x}$ are used, and rectangle edges are
identified where zero crossings occur.  In the case of more than one
zero crossing, the crossing with largest rate of change in $\Delta$ is chosen for
the partition; if two crossings have the same magnitude, the one
closest to the rectangle center is chosen to prevent thin rectangles
which reduce efficiency.  If none of the above partitioning criteria
are met and the efficiency ratio still has not satisfied the
$\varepsilon_{\mathrm{min}}$ value, the rectangle is bisected along
its longest dimension.

\subsection{Coarse to fine interpolation\label{sec:cfi}}
High-order interpolation of the distribution function from a coarser
level to a finer level is performed (1) to calculate cell-averaged
values in ghost-cells, (2) to interpolate charge and current densities
to the finest level of the grid and (3) to interpolate data to a
refined cell after performing mesh adaption. For high-order
discretizations, which are needed (for example) to treat the disparate scales of
the discretization in an adaptive solver, low order slope limited
interpolation is not suitable.  Instead methods such as filtered high
order interpolations~\cite{Ray2007}, Weighted Essentially
Non-Oscillatory (WENO) techniques~\cite{Shu1997,Sebastian2003,Shen2011} and least squares
methods~\cite{Mccorquodale2011} can be used.

In this work, we use a simple fourth-order conservative least square
interpolation method. Coarse to fine interpolation of
the distribution function, i.e.\ interpolation from
$\tilde f_{k,l}^{i,i}$ to $\tilde f_{rk+m,rl+n}^{i+1,i+1}${,} where
$m,n=0,\dots,r-1${,} is performed in two steps, by first interpolating
$\tilde f_{k,l}^{i,i}$ to $\tilde f_{rk+m,l}^{i+1,i}$ and then { to}
$\tilde f_{rk+m,rl+n}^{i+1,i+1}$.  Hence, it is sufficient to describe
only the interpolation step from $\tilde f_{k,l}^{i,i}$ to
$\tilde f_{rk+m,l}^{i+1,i}$, because the interpolation in the other coordinate is analogous.  {This is performed by first introducing
\begin{equation}
   h(x)=\frac{1}{\Delta p_{x,i}}\int_{p_{x,l-\frac{1}{2}}^{i}}^{p_{x,l+\frac{1}{2}}^{i}}f(x,p_x)\, \mathrm{d}p_x
\end{equation}
which is interpolated using a third degree polynomial $q(x)=\sum_{j=0}^3a_jx^j$.
Defining $\cramped{I(a,b)=1/(b-a)\int_a^b q(x)\,\mathrm{d}x}$, $q(x)$
is  determined by the least squares solution to
$\tilde
  f^{i,i}_{k+m,l}=I(x_{k+m-\frac{1}{2}}^{i},x_{k+m+\frac{1}{2}}^{i})$
for $m=\pm 1,\pm 2$ under the condition
$\tilde f^{i,i}_{k,l} =
  I(x_{k-\frac{1}{2}}^{i},x_{k+\frac{1}{2}}^{i})$,
ensuring { particle number} conservation.  The values for
$\tilde f_{rk+m,l}^{i+1,i}$ are calculated from
$I(x_{rk+m-\frac{1}{2}}^{i+1},x_{rk+m+\frac{1}{2}}^{i+1})$,
where $x_{rk+m\pm\frac{1}{2}}^{i+1}$ are the end-points of
the cells in the refined level.

\subsection{Discretization of the Vlasov equation\label{sec:vlasov}}
We introduce a finite volume method which decomposes the
discretized advection operator into fluxes across cell
boundaries.
On each level $i$ in the mesh, the Vlasov equation is averaged over cells
\begin{equation}
   \Delta_{k,l}^{i}=[x_{k-\frac{1}{2}}^{i},x_{k+\frac{1}{2}}^{i}]\times[p_{x,l-\frac{1}{2}}^{i},p_{x,l+\frac{1}{2}}^{i}]
\end{equation}
in the mesh, resulting in a set of ordinary differential equations (ODEs):
\begin{align}
\frac{\mathrm{d} \tilde f^{i,i}_{k,l}}{\mathrm{d}t} = &-\frac{1}{\Delta x_i}\left(\langle F_x f \rangle_{k+\frac{1}{2},l}^{i}-\langle F_x f \rangle_{k-\frac{1}{2},l}^{i}\right) \nonumber\\
&-\frac{1}{\Delta p_{x,i}}\left(\langle F_{p_x} f \rangle_{k,l+\frac{1}{2}}^{i}-\langle F_{p_x} f \rangle_{k,l-\frac{1}{2}}^{i}\right) \label{eq:fvm}
\end{align}
where $\cramped{\tilde f^{i,i}_{k,l}}$ is the cell-averaged value of the
distribution function, and $\langle
F_xf\rangle_{k\pm\frac{1}{2},l}^{i}$, $\langle
F_{p_x}f\rangle_{k,l\pm\frac{1}{2}}^{i}$ denote fluxes. The fluxes are
defined as:
\begin{align}
\MoveEqLeft \langle F_x f \rangle_{k+\frac{1}{2},l}^{i}= \nonumber \\
&\frac{1}{\Delta p_{x,i}}\int_{p_{x,l-\frac{1}{2}}^{i}}^{p_{x,l+\frac{1}{2}}^{i}}F_x(x_{k+\frac{1}{2}}^{i},p_x)f(x_{k+\frac{1}{2}}^{i},p_x)\,\mathrm{d}p_x
\end{align}
and
\begin{align}
\MoveEqLeft \langle F_{p_x} f \rangle_{k,l+\frac{1}{2}}^{i}= \nonumber \\
&\frac{1}{\Delta x_{i}}\int_{x_{k-\frac{1}{2}}^{i}}^{x_{k+\frac{1}{2}}^{i}}F_{p_x}(x,p_{x,l+\frac{1}{2}}^{i})f(x,p_{x,l+\frac{1}{2}}^{i})\,\mathrm{d}x,
\end{align}
where
\begin{equation*}
F_x= \frac{p_x}{m\gamma}\quad\text{and}\quad  F_{p_x}= q\left[E_x+\frac{1}{m\gamma}(\bm{p}\times \bm{B})_x\right].
\end{equation*}

The ODEs \eqref{eq:fvm} are exact and the numerical approximations enter in
the calculation of the flux terms
$\langle F_xf\rangle_{k+\frac{1}{2},l}^{i}$ and
$\langle F_{p_x}f\rangle_{k,l+\frac{1}{2}}^{i}$.
Despite the gain from the fast convergence of
{high-order methods { in the calculation of the flux terms, high-order methods suffer from spurious oscillations in regions with under-resolved gradients that could trigger numerical instabilities.}}
To avoid this, we
use a Flux Corrected Transport (FCT) algorithm in the calculation
of the advective terms in the Vlasov equation  as suggested in Ref.\ \cite{Banks2010}.
The implementation of
FCT, which is described in \ref{sec:timestep}, mixes  low and high-order fluxes in such a way as to maximize
the high-order flux without introducing unphysical properties in the
distribution function. Whilst investigations of modified  and refined FCT methods to limit spurious oscillations associated with high-order methods are underway \cite{Chaplin2015}, we use the original implementation in Ref. \cite{ZALESAK}. The reduction in order of accuracy for the limited solution in regions with under-resolved gradients is compensated by the use of a finer grid in exactly these regions.

We evaluate each flux
in two different ways, using a low and a high-order method,
respectively. The low order fluxes, denoted
$\langle F_xf\rangle_{k+\frac{1}{2},l}^{i,L}$,
$\langle F_{p_x}f\rangle_{k,l+\frac{1}{2}}^{i,L}$, are evaluated using
first order upwinding, i.e.\ as the product of the face-averaged force
terms $\langle F_x\rangle_{k+\frac{1}{2},l}^{i}$ or
$\langle F_{p_x}\rangle_{k,l+\frac{1}{2}}^{i}$ and the cell-averaged
value of the distribution function in the upwind
cell.
In section \ref{sec:timestep}, the advection operator is calculated by
blending these with high-order fluxes
$\cramped{\langle F_xf\rangle_{k+\frac{1}{2},l}^{i,H}}$,
$\cramped{\langle F_{p_x}f\rangle_{k,l+\frac{1}{2}}^{i,H}}$, using the FCT
algorithm, to obtain a stable scheme which does not create unphysical
extrema in the distribution function.

To obtain a second order accurate calculation of $\cramped{\langle
    F_xf\rangle_{k+\frac{1}{2},l}^{i,H}}$ and $\cramped{\langle
F_{p_x}f\rangle_{k,l+\frac{1}{2}}^{i,H}}$, the cell-averaged and
face-averaged quantities can be approximated by their cell and face
centered values. However, for a finite volume scheme beyond second
order accuracy it is necessary to distinguish cell-centered and
averaged values as well as using accurate quadrature rules for the
face-integrals. Here, we follow Ref.~\cite{Zhang2012} and add
corrections to the midpoint approximation of the face-integrals using
the transverse derivatives, yielding:
\begin{align}
\MoveEqLeft[1.5] \langle F_xf \rangle_{k+\frac{1}{2},l}^{i,H} = \nonumber\\
&\langle F_x\rangle_{k+\frac{1}{2},l}^{i}\langle f\rangle_{k+\frac{1}{2},l}^{i}+\frac{1}{48}(\langle F_x\rangle_{k+\frac{1}{2},l+1}^{i}-\langle F_x\rangle_{k+\frac{1}{2},l-1}^{i}) \nonumber \\
&\cdot(\langle f\rangle_{k+\frac{1}{2},l+1}-\langle f\rangle_{k+\frac{1}{2},l-1})
\end{align}
and
\begin{align}
\MoveEqLeft[1.5] \langle F_{p_x}f \rangle_{k,l+\frac{1}{2}}^{i,H} = \nonumber \\
&\langle F_{p_x}\rangle_{k,l+\frac{1}{2}}\langle f\rangle_{k,l+\frac{1}{2}}^{i}+\frac{1}{48}(\langle F_{p_x}\rangle_{k+1,l+\frac{1}{2}}^{i}-\langle F_{p_x}\rangle_{k-1,l+\frac{1}{2}}^{i}) \nonumber\\
&\cdot(\langle f\rangle_{k+1,l+\frac{1}{2}}^{i}-\langle f\rangle_{k-1,l+\frac{1}{2}}^{i})
\end{align}
respectively. This is a fourth-order accurate expression for the
flux-integrals. Here, the face-averaged values of $F_x$ and $F_{p_x}$ are defined as
\begin{align}
\langle F_x  \rangle_{k+\frac{1}{2},l}^{i}&=\frac{1}{\Delta p_{x,i}}\int_{p_{x,l-\frac{1}{2}}^{i}}^{p_{x,l+\frac{1}{2}}^{i}}F_x(x_{k+\frac{1}{2}}^{i},p_x)\,\mathrm{d}p_x, \\
\langle F_{p_x} \rangle_{k,l+\frac{1}{2}}^{i}&=\frac{1}{\Delta x_{i}}\int_{x_{k-\frac{1}{2}}^{i}}^{x_{k+\frac{1}{2}}^{i}}F_{p_x}(x,p_{x,l+\frac{1}{2}}^{i})\,\mathrm{d}x,
\end{align}
and the distribution functions
\begin{align}
\langle f \rangle_{k+\frac{1}{2},l}^{i}&=\frac{1}{\Delta p_{x,i}}\int_{p_{x,l-\frac{1}{2}}^{i}}^{p_{x,l+\frac{1}{2}}^{i}}f(x_{k+\frac{1}{2}}^{i},p_x)\,\mathrm{d}p_x,\\
\langle f \rangle_{k,l+\frac{1}{2}}^{i}&=\frac{1}{\Delta x_{i}}\int_{x_{k-\frac{1}{2}}^{i}}^{x_{k+\frac{1}{2}}^{i}}f(x,p_{x,l+\frac{1}{2}}^{i})\,\mathrm{d}x,
\end{align}
which are used to evaluate the fluxes.

The Hamiltonian for a particle is
\begin{equation}
H=mc^2\gamma(p_x,\bm{A}_\perp(x))+q\phi(x).
\end{equation}
Using Hamilton's equations $\partial H/\partial p_x=F_x$ and \linebreak
$-\partial H/\partial x = F_{p_x}$, the face-averaged force terms take
the forms
\begin{align}\label{eq:fx}
\langle F_x\rangle_{k+\frac{1}{2},l}^{i}=&\frac{mc^2}{\Delta p_{x,i}}\left[\gamma\left(p_{x,l+\frac{1}{2}}^{i},\bm{A}_\perp\left(x^{i}_{k+\frac{1}{2}}\right)\right)\right. \nonumber\\
&-\left.\gamma\left(p_{x,{l-\frac{1}{2}}}^{i},\bm{A}_\perp\left(x^{i}_{k+\frac{1}{2}}\right)\right)\right]
\end{align}
and
\begin{align}\label{eq:fpx}
\langle F_{p_x}\rangle_{k,l+\frac{1}{2}}^{i}=& q\tilde E_{x,{k}}-\frac{mc^2}{\Delta x_i}\left[\gamma\left(p_{x,l+\frac{1}{2}}^{i},\bm{A}_\perp\left(x^{i}_{k+\frac{1}{2}}\right)\right)\right. \nonumber \\
&-\left.\gamma\left(p_{x,l+\frac{1}{2}}^{i},\bm{A}_\perp\left(x^{i}_{k-\frac{1}{2}}\right)\right)\right].
\end{align}
This indicates that the vector potential must be evaluated on the
spatial faces.  For the electric field, we can directly evaluate its
cell-averaged value through the electrostatic potential for which we
solve.

To evaluate $\langle f\rangle_{k+\frac{1}{2},l}^{i}$ (and similarly
for $\langle f\rangle_{k,l+\frac{1}{2}}^{i}$) an upwind biased
WENO-type reconstruction is used \cite{Shu1997,Banks2010}. The WENO
scheme makes use of a four cell stencil involving the nearest
and next nearest neighbours in the normal direction of the
$(k+\frac{1}{2})$-face. These four cells are divided into two
sub-stencils, defining two third order interpolants
\begin{align}
   p_L&=(-\tilde f_{k-1,l}^{i,i}+5\tilde f_{k,l}^{i,i}+2\tilde f_{k+1,l}^{i,i})/6\\
   \shortintertext{and}
   p_R&=(2\tilde f_{k,l}^{i,i}+5\tilde f_{k+1,l}^{i,i}-\tilde f_{k+2,l}^{i,i})/6.
\end{align}
A weighted interpolant is then obtained by setting
\begin{equation}
   \langle f\rangle_{k+\frac{1}{2},l}^{i}=\beta_Lp_L+\beta_Rp_R,
\end{equation}
where $\beta_L+\beta_R=1$. The values $\beta_L=\beta_R=1/2$ result in a
fourth order central difference approximation of $\langle
f\rangle_{k+\frac{1}{2},l}^{i}$.

Following the WENO procedure, $\beta_L$, $\beta_R$ are calculated
based on estimates for the smoothness of the interpolant and approach
$\beta_L=\beta_R=1/2$ in the limit of a smooth distribution
function. However, in contrast to the conventional WENO algorithm
\cite{Shu1997}, we follow Ref.~\cite{Banks2010}, so that the largest
weight is assigned to the value of $p_L$ or {$p_R$} that is associated
with the upwind stencil.

Up to this point, we have not considered the effect of
coarse-fine interfaces in the calculation of fluxes. Coarse to fine
interfaces are illustrated in Figure \ref{fig:cellidx}. At the cell
faces which constitute a coarse-fine interface, i.e.\ the exterior
boundary of a fine level $L_{i+1}$, fluxes on the fine level
and coarser level $L_i$ must be defined consistently in order to
obtain conservation of particle number. Conservation up to machine
error is critical due to the feedback of the charge density through
the longitudinal electric field. For a spatial face $(k+{1}/{2},l)$ in
level $L_{i}$, the flux is defined using the fluxes calculated on the
$r$ constituent finer cell faces:
\begin{equation}
   \langle F_xf\rangle_{k+\frac{1}{2},l}^{i}=1/r\sum_{m=0}^{r-1}\langle F_xf\rangle_{(r+1)k-\frac{1}{2},l+m}^{i+1}.
\end{equation}
This enforces that the number of particles flowing from the fine cells is the same as the
number of particles that enter the coarse cell. Faces associated with
the momentum coordinate are treated similarly.

\subsection{Evaluation of charge and current densities}

To evaluate a general moment
\begin{equation}
M(x)=\int m(x,p_x)\,\mathrm{d}p_x{,}
\end{equation}
a reduction operation is performed over the different levels and
rectangles, yielding a sum of the form:
\begin{equation}
M(x)=\sum_{L_i}\sum_l \chi(x,l,i)I(x,l,i)\label{eq:moment}.
\end{equation}
Here, $\chi(x,l,i)$ is equal to one if $i$ is the finest
level containing $\{x\}\times [p_{x,l-\frac{1}{2}}^{i},p_{x,l+\frac{1}{2}}^{i}]$ or is otherwise zero, and $ I(x,l,i)$ approximates
\begin{equation}
   \int_{p_{x,l-\frac{1}{2}}^{i}}^{p_{x,l+\frac{1}{2}}^{i}}m(x,p_x)\,\mathrm{d}p_x,
\end{equation}
using solution quantities at the level $L_i$.

To evaluate the cell-averaged charge density, values of the
distribution function, represented on different levels $i$, are
interpolated using least-square interpolation to the finest spatial
level $n_\text{max}$. The interpolated values are denoted $\tilde
f_{k,l}^{n_\text{max},i}$ and the charge density is then calculated by
the reduction:
\begin{equation}
\tilde \rho_k=\sum_{L_i}\sum_l \Delta p_{x,i}\chi(k,l,i)\tilde f_{k,l}^{n_\text{max},i}.\label{eq:rho}
\end{equation}
The only source of error introduced in this relation is due to
the interpolation to the finest grid with respect to the spatial
coordinate.

Accurate calculation of the current density is more challenging as the
integrand in Eq.~\eqref{eq:current} has a more complicated dependence on
$x$ and $p_x$ (through $\gamma(p_x,\bm{A}_\perp (x))$, $f(x,p_x)$ and
$\bm{A}_\perp (x)$), but is of critical importance for the interaction
between the plasma and the electromagnetic field.  Using the interpolated
values $\tilde f_{k,l}^{n_\text{max},i}$, the spatial cell averaging
is made to second-order accuracy by commuting the cell averaging and
evaluation of the integrand with respect to the spatial coordinate:
\begin{align}\nonumber
\tilde{\bm{J}}_k&=-q\tilde{\bm{A}}_{\perp,k}\int \frac{1}{\gamma(p_x,\tilde{\bm{A}}_{\perp,k})}\tilde f^{n_\text{max}}_k(p_x)\,\mathrm{d}p_x\\
&:=-q\tilde{\bm{A}}_{\perp,k}\tilde L_k
\end{align}
As for the charge density, $\tilde L_k$ is evaluated using
Eq.~\eqref{eq:moment}, but in contrast to Eq.~\eqref{eq:rho}, the integrand
$I(\tilde{\bm{A}}_{\perp,k},l,i)$ needs further consideration and use
of a quadrature formula based on cell averaged quantities.  The
situation is similar to the flux calculation in \ref{sec:vlasov} and
the fourth order scheme:
\begin{align}\label{eq:cur}
\MoveEqLeft \int_{p_{x,l-\frac{1}{2}}^{i}}^{p_{x,l+\frac{1}{2}}^{i}}\frac{1}{\gamma(p_x,\tilde{\bm{A}}_{\perp,k})}\tilde f^{n_\text{max}}_k(p_x)\,\mathrm{d}p_x= \nonumber\\
&\frac{\Delta p_{x,i}}{\tilde \gamma^{n_\text{max},i}_{k,l}(\tilde{\bm{A}}_{\perp,k})}\tilde f^{n_\text{max},i}_{k,l} \nonumber\\
&+\frac{\Delta p_{x,i}}{48}\parens[\Bigg]{\frac{1}{\tilde \gamma^{n_\text{max},i}_{k,l+1}(\tilde{\bm{A}}_{\perp,k})}-\frac{1}{\tilde \gamma^{n_\text{max},i}_{k,l-1}(\tilde{\bm{A}}_{\perp,k})}} \nonumber\\
&\times\parens[\Bigg]{\tilde f^{n_\text{max},i}_{k,l+1}-\tilde f^{n_\text{max},i}_{k,l-1}}
\end{align}
is used. In this expression, the momentum average values for
$1/\gamma$ are calculated analytically:
\begin{align}\label{eq:gamma}
\MoveEqLeft \frac{1}{\tilde \gamma^{n_\text{max},i}_{k,l}(\tilde{\bm{A}}_{\perp,k})} \nonumber\\
&=\frac{1}{\Delta p_{x,i}}\int_{p_{x,l-\frac{1}{2}}^{i}}^{p_{x,l+\frac{1}{2}}^{i}}\frac{1}{\gamma(p_x,\tilde{\bm{A}}_{\perp,k})}\,\mathrm{d}p_x\\
&=\frac{mc}{\Delta p_{x,i}}\log\left(\frac{\gamma(p^{i}_{l+\frac{1}{2}},\tilde{\bm{A}}_{\perp,k})+p^{i}_{l+\frac{1}{2}}/mc}{\gamma(p^{i}_{l-\frac{1}{2}},\tilde{\bm{A}}_{\perp,k})+p^{i}_{l-\frac{1}{2}}/mc}\right).
\end{align}

\subsection{Discretization of Maxwell's equations\label{sec:maxwell}}

To solve the Vlasov-Maxwell system self-consistently, we perform a
spatial discretization of the equations for the transverse fields,
resulting in a set of ODEs which are simultaneously time stepped with
the kinetic equation.  We discretize $E_y$, $E_z$, $A_y$ and $A_z$
using the cell-centred values $\tilde E_{y,k}$, $\tilde E_{z,k}$,
$\tilde A_{y,k}$ and $\tilde A_{z,k}$, i.e.\ averages over the same
spatial cells as for the kinetic equation.  To avoid odd-even
decoupling, we represent $B_y$ and $B_z$ as cell-centred averages
$\tilde B_{y,k+\frac{1}{2}}$ and $\tilde B_{z,k+\frac{1}{2}}$, at a
staggered grid.  The fields are discretized on the finest level for
the distribution function. This is motivated by the fact that
Maxwell's equations only depend on the spatial coordinate making the
computation cheap compared to the computation for the Vlasov
equations, in spite of the use of a fine mesh.

The equations for the transverse fields are cell averaged and the
spatial derivatives are discretized to fourth order, yielding
\begin{align}\label{eq:t1}
   \frac{\partial \tilde B_{y,k+\frac{1}{2}}}{\partial t}&=\frac{ -\tilde E_{z,k+2}+27\tilde E_{z,k+1}-27\tilde E_{z,k}+\tilde E_{z,k-1}}{24\Delta x_{n_\text{max}}}, \\ \label{eq:t2}
   \frac{\partial \tilde B_{z,k+\frac{1}{2}}}{\partial t}&=-\frac{ -\tilde E_{y,k+2}+27\tilde E_{y,k+1}-27\tilde E_{y,k}+\tilde E_{y,k-1}}{24\Delta x_{n_\text{max}}},\\ \label{eq:t3}
   \epsilon_0\mu_0\frac{\partial\tilde E_{y,k}}{\partial t}&= \nonumber\\
   \MoveEqLeft[3] -\frac{ -\tilde B_{z,k+\frac{3}{2}}+27\tilde B_{z,k+\frac{1}{2}}-27\tilde B_{z,k-\frac{1}{2}}+\tilde B_{z,k-\frac{3}{2}}}{24\Delta x_{n_\text{max}}}-\mu_0\tilde J_{y,k},\\ \label{eq:t4}
   \epsilon_0\mu_0\frac{\partial\tilde E_{z,k}}{\partial t}&=\nonumber\\
   \MoveEqLeft\frac{ -\tilde B_{y,k+\frac{3}{2}}+27\tilde B_{y,k+\frac{1}{2}}-27\tilde B_{y,k-\frac{1}{2}}+\tilde B_{y,i-\frac{3}{2}}}{24\Delta x_{n_\text{max}}}-\mu_0\tilde J_{z,k},
\end{align}
and
\begin{equation}
   \frac{\partial\tilde A_{y,k}}{\partial t}=-\tilde E_{y,k}, \qquad \frac{\partial\tilde A_{z,k}}{\partial t}=-\tilde E_{z,k}.\label{eq:t5}
\end{equation}
Here, $\tilde J_k$ denotes the cell-averaged current density. In
simulations of laser matter interaction, a laser pulse is implemented
as a Dirichlet boundary condition at the left hand side of the
simulation box.

Regarding the use of the transverse fields to evaluate the
coefficients in the Vlasov equation, we note that Eq.~\eqref{eq:fx}
and Eq.~\eqref{eq:fpx} depend on the
point values of the transverse vector potential at
$x_{k+\frac{1}{2}}^{n_\text{max}}$. To calculate the vector potential
on the spatial faces from its cell-averaged values, we use a fourth
order WENO interpolation, similar to the one that was used for face
interpolation of the distribution function, but in this case without
upwind biasing of the smoothness indicators. Including a
non-linear scheme here is primarily done for the sake of safety: to squelch
numerical sources of noise. Being a one dimensional interpolation, the
computational overhead is minimal.

For the equation with the longitudinal electric field $E_x$, we
introduce $\phi$, such that $E_x=-\partial\phi/\partial x$. The
potential $\phi$ satisfies the Poisson equation $\Delta
\phi=-\rho/\epsilon_0$. A fourth order discretization is given by
Ref.~\cite{Hittinger2013}:
\begin{equation}
\frac{30\tilde\phi_k-{16}(\tilde\phi_{k+1}+\tilde\phi_{k-1})+{}(\tilde\phi_{k+2}+\tilde\phi_{k-2})}{12\Delta x_{n_\text{max}}^2}=\tilde\rho_k/\epsilon_0\label{eq:poi}
\end{equation}
where $\tilde\phi_k$ is the cell-averaged potential and $\tilde\rho_k$
is the cell-averaged charge density.

For simulations of laser matter interaction, where no charge leaves
the simulation box, we take the electric field at both boundaries of
the simulation box to be
zero.
The boundary conditions are implemented by splitting the potential in
two parts $\phi=\phi_1+\phi_2$ where $\phi_1$ satisfies
Eq.~\eqref{eq:poi} with periodic boundary conditions and hence makes
it possible to avoid modification of the discretization at the
boundary points. Furthermore, we take $\phi_2=-E_0x$, which satisfies
the homogeneous Poisson equation and choose $E_0$ such that the
homogeneous Neumann condition is fulfilled at the right boundary.

The cell-averaged electric field, which is used to evaluate $F_{p_x}$,
is then obtained to fourth order from:
\begin{align}
\MoveEqLeft[1] \tilde E_{x,k}=\nonumber\\
&-\frac{1}{12\Delta x_{n_\text{max}}}\left[8(\tilde \phi_{k+1}-\tilde\phi_{k-1})-\tilde \phi_{k+2}+\tilde \phi_{k-2}\right]+E_0.
\end{align}

\subsection{Time advancement\label{sec:timestep}}

For time advancement of the ODEs in the discretization of the
Vlasov-Maxwell system we use the Runge-Kutta method ARK4(3)6L[2]SA,
which is a six stage, fourth-order accurate method \cite{Kennedy}. It
is a suitable choice for future extensions involving a Fokker-Planck
diffusion operator.  In such extensions the advection operator would
be treated explicitly and the diffusion term implicitly
\cite{Zhang2012}.

The combined set of ODEs for the transverse electromagnetic fields and
the distribution function can be regarded as a system of equations of
the form
\begin{equation}
   \frac{\partial \bm{F}(t)}{\partial t}=\bm{L}(\bm{F}),
\end{equation}
where $\bm{L}$ stands for the respective right hand sides of Eqs. \eqref{eq:fvm}, \eqref{eq:t1}-\eqref{eq:t5}. Denoting the
solution at time step $t_n$ by $\bm{F}(t_n)$, a sequence of solutions,
$\bm{F}^i(t_n^i)=\bm{F}(t_n)+\Delta t_n \bm{L}_i$, at intermediate times
$t_n^i=c_i\Delta t+t_n$ are constructed, for $i=1,\dots,6$.
{ Here}, $\bm{L}_i$ is a linear combination of $\bm{L}(\bm{F})$
at the previous intermediate time steps:
\begin{equation}
\bm{L}_i=\sum_{j=1}^{i-1} a_{ij}\bm{L}\left(\bm{F}^j(t^j_n)\right).
\end{equation}
Once $\bm{L}\left(\bm{F}^i(t_{n}^i)\right)$ has been calculated for
$i=1,\dots,6$, the solution at time step $t_{n+1}=t_n+\Delta t$ is
calculated from
\begin{equation}
\bm{F}(t_{n+1})=\bm{F}(t_{n})+\Delta t\sum_{j=1}^6b_j\bm{L}\left(\bm{F}^j(t_{n}^j)\right),
\end{equation}
where the values of $a_{ij}$, $b_{j}$ and $c_i$ can be found in Ref.~\cite{Kennedy}.

The part of $\bm{L}(\bm{F})$ that corresponds to Maxwell's equations
is evaluated using the discretization in subsection
\ref{sec:maxwell}. On the other hand, for the Vlasov part, the FCT
algorithm is applied in order to limit anti-diffusive fluxes, which may
cause unphysical extrema and is the reason behind introducing the low-order
$\cramped{\left(\langle F_xf\rangle_{k+\frac{1}{2},l}^{i,L}\right.}$, $\cramped{\left.\langle
F_{p_x}f\rangle_{k,l+\frac{1}{2}}^{i,L}\right)}$ and high-order $\cramped{\left(\langle
F_xf\rangle_{k+\frac{1}{2},l}^{i,H}\right.}$, $\cramped{\left.\langle
F_{p_x}f\rangle_{k,l+\frac{1}{2}}^{i,H}\right)}$ fluxes in subsection
\ref{sec:vlasov}. Although the  FCT algorithm is found to be necessary to obtain a stable scheme, its use comes at the expense of introducing a source of hyper-diffusivity in regions with under-resolved gradients. To mitigate this, the criterion for refinement in subsection \ref{sec:error}  involves terms proportional to derivatives of the distribution function.

When applying the Runge-Kutta method, the
flux that is used to calculate the solution at an intermediate or
final state in the time stepping algorithm is a linear combination of
the fluxes calculated from the intermediate solutions at earlier
stages. The generic structure of an advancement of the distribution
function takes the form
\begin{align}
f_{k,l}^1 =\;& f_{k,l}^0-F_{x}(k+{1}/{2},l)+F_{x}(k-1/2,l)\nonumber\\
&-F_{p_x}(k,l+{1}/{2})+F_{p_x}(k,l-{1}/{2}),
\end{align}
where $f_{k,l}^0$, $f_{k,l}^1$ is the solution before and after time
advancement, respectively, and $F_x$, $F_{p_x}$ denote linear
combinations of flux terms calculated from intermediate
solutions. Here, we have simplified the notation by ignoring the level
in the mesh, on which the quantities are defined.  The combined flux
terms $F_x$, $F_{p_x}$ can  be evaluated using either the low or high
order fluxes, denoted $F_x^L$, $F_{p_x}^L$ and $F_x^H$, $F_{p_x}^H$,
respectively. For the high order flux, the weights of the fluxes at intermediate steps are those according to the Runge-Kutta method. However, to ensure that the low order solution is positive, the upwind flux at time $t_n$ is used.

Following Ref.~\cite{ZALESAK},
we first calculate an approximation to $f_{k,l}^1$ using the low order
fluxes
\begin{align}
f_{k,l}^{1,L} =\;& f_{k,l}^0-F_{x}^L(k+{1}/{2},l)+F_{x}^L(k-1/2,l) \nonumber\\
&-F_{p_x}^L(k,l+{1}/{2})+F_{p_x}^L(k,l-{1}/{2}).
\end{align}
This is followed by defining anti-diffusive fluxes
\begin{align}
   A_{k+\frac{1}{2},l}&=F_{x}^H(k+{1}/{2},l)-F_{x}^L(k+{1}/{2},l) \\
   \shortintertext{and}
   A_{k,l+\frac{1}{2}}&=F_{p_x}^H(k,l+1/2)-F_{p_x}^L(k,l+1/2).
\end{align}
Based on these, the total amount of anti-diffusive fluxes into ($P^+_{k,l}$) and out ($P^-_{k,l}$) of a cell is defined by
\begin{align}
P^+_{k,l}=\;&\max(0,A_{k-\frac{1}{2},l})-\min(0,A_{k+\frac{1}{2},l})\nonumber\\
&+\max(0,A_{k,l-\frac{1}{2}})-\min(0,A_{k,l+\frac{1}{2}})
\shortintertext{and}
P^-_{k,l}=\;&\max(0,A_{k+\frac{1}{2},l})-\min(0,A_{k-\frac{1}{2},l})\nonumber\\
&+\max(0,A_{k,l+\frac{1}{2}})-\min(0,A_{k,l-\frac{1}{2}}).
\end{align}
Furthermore, \red{for cells that share faces with the cell with indices
$k$ and $l$, we define $f^{\text{max}}_{k,l}$ and $f^{\text{min}}_{k,l}$
to be the maximum and minimum values, respectively, of both $f_{k,l}^{1,L}$ and
$f_{k,l}^{0}$.} The maxima and minima are used to define
$Q_{k,l}^+=f_{k,l}^{\text{max}}-f_{k,l}^{1,L}$ and
$Q_{k,l}^-=f_{k,l}^{1,L} -f_{k,l}^{\text{min}} $, which describe the
maximum allowed increase/decrease that is compatible { { with not} creating unphysical extrema}. Based on this, a maximum fraction of the anti-diffusive
flux that can be allowed to enter/exit a cell is given by
\begin{equation*}
R^+_{k,l}=\min(1,Q_{k,l}^+/P_{k,l}^+) \quad \text{and}\quad\ R^-_{k,l}=\min(1,Q_{k,l}^-/P_{k,l}^-).
\end{equation*}
Depending on the sign of the anti-diffusive flux across the different faces, the
fraction of the anti-diffusive flux that can be admitted is given by
\begin{equation}
C_{k+\frac{1}{2},l}=
\begin{dcases}
   \min(R^+_{k+1,l},R^-_{k,l})&\text{if}\; A_{k+\frac{1}{2},l}\geq 0\\
   \min(R^+_{k,l},R^-_{k+1,l})&\text{if}\; A_{k+\frac{1}{2},l}< 0
\end{dcases}
\end{equation}
and
\begin{equation}
C_{k,l+\frac{1}{2}}=
\begin{dcases}
   \min(R^+_{k,l+1},R^-_{k,l})&\text{if}\; A_{k,l+\frac{1}{2}}\geq 0\\
   \min(R^+_{k,l},R^-_{k,l+1})&\text{if}\; A_{k,l+\frac{1}{2}}< 0.
\end{dcases}
\end{equation}
These fractions are finally used to add the maximum allowed contribution from the higher order fluxes
\begin{align}
f_{k,l}^1 =\;& f_{k,l}^{1,L}-C_{k+\frac{1}{2},l}A_{k+\frac{1}{2},l}+C_{k-\frac{1}{2},k}A_{k-\frac{1}{2},k}\nonumber\\
&-C_{k,l+\frac{1}{2}}A_{k,l+\frac{1}{2}}+C_{k,l-\frac{1}{2}}A_{k,l-\frac{1}{2}}.
\end{align}

For an explicit time-stepping scheme to be stable, the time step must fulfill a Courant-Friedrichs-Lewy (CFL) condition.
Here, the time step is determined by enforcing a user provided maximum CFL number such that the worst case coefficients (which scale with $a_0$) in the simulation do not exceed this value. The available range of CFL numbers is dictated by the stability region of the Runge-Kutta algorithm.

\subsection{Refinement indicator\label{sec:error}}

To concentrate finer resolution to parts of phase-space where the
distribution function has a complicated structure, we choose a refinement indicator
which depends on the magnitude of the distribution function as well as
 its first and second derivatives:
\begin{align}
\delta=\;&w_{x,1}|\tilde f_{k+1,l}^{i,i}-\tilde f_{k-1,l}^{i,i}|+w_{p_x,1}|\tilde f_{k,l+1}^{i,i}-\tilde f_{k,l-1}^{i,i}|\nonumber\\
&+w_{x,2}|\tilde f_{k+1,l}^{i,i}-2\tilde f_{k,l}^{i,i}+\tilde f_{k-1,l}^{i,i}|+w_{p_x,2}|\tilde f_{k,l+1}^{i,i} \nonumber\\
&-2\tilde f_{k,l}^{i,i}+\tilde f_{k,l-1}^{i,i}|+w_f \tilde f_{k,l}^{i,i}.
\end{align}
Here, the constants $w_{x,1}r^{-i/2}$, $w_{p_x,1}r^{-i/2}$, $w_{x,2}r^{-2i}$,\linebreak
$w_{p_x,2}r^{-2i}$, $w_f r^{-i}$ are chosen such that they are all
inversely proportional to $f_{\text{max}}$, where $f_\text{max}$ is
the maximum value of the distribution function, {which} is determined from the initial condition.

In addition to the criteria for mesh refinement, we have implemented
the possibility to specify a minimum level of refinement as a function
of the phase space coordinates. This makes it possible to guarantee
that parts of phase-space which are known to carry important
information about the distribution function, and have complex dynamics
{\it a priori}, are sufficiently resolved. The plasma-vacuum interface
is an example of such a region, where most of the laser-plasma
interaction occurs.

\section{Numerical benchmarking}

To benchmark {\sc veritas}, we present results from Vlasov-Maxwell
simulations in two cases.  In the first case we consider a situation
with a circularly polarized (CP) laser pulse impinging on an overdense
electron plasma.  In this case, it is possible to derive analytical
solutions for the electron density by using fluid theory, and these
analytical results in turn can be compared with numerical results of
{\sc veritas}.

The second benchmarking example will be a study of laser plasma
interaction in three different regimes of laser-plasma interaction.
Here we will
make a comparison with results of the well-established PIC code {\sc
 epoch}~\cite{arber2015}.

\begin{figure}[tb]
    \includegraphics[width=\columnwidth]{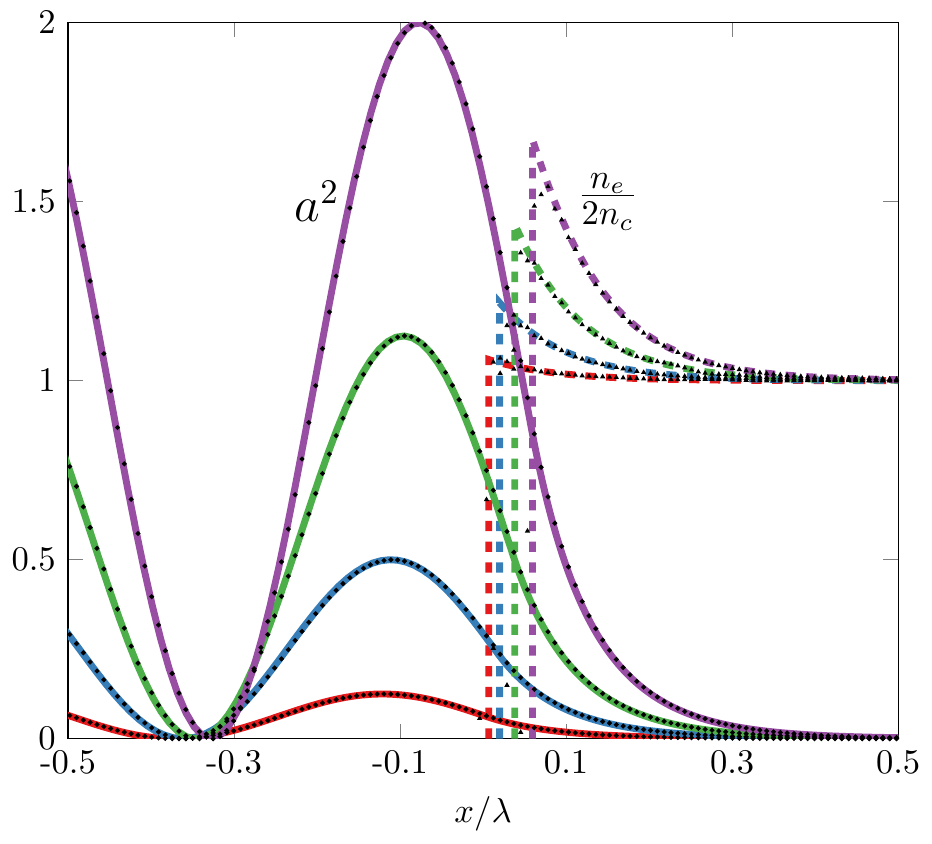}
\caption{\label{fig:quasistationary}Comparison of analytical solutions
  for density (dashed lines) and squared normalized vector potential
  (solid lines) and solutions that are calculated using {\sc veritas}
  (black markers), for a circularly polarized laser pulse that
  impinges on an overdense plasma. The density has been normalized by
  $2n_c$, to fit on the same scale as the squared vector
  potential. The analytical results for different intensities are
  labelled with the colours red ($a_0=0.25$), blue ($a_0=0.50$), green
  ($a_0=0.75$) and purple ($a_0=1.00$).  }
\end{figure}

\subsection{Circularly polarized light impinging on a plasma slab -- comparison to  analytic theory\label{sec:benchmark1}}

We study a circularly polarized laser pulse interacting
with an overdense plasma, i.e.\ a plasma with a density $n_0$ higher than the critical
density $n_c=\omega^2m_e\epsilon_0/e^2$, where $\omega$ is the laser frequency.
In particular, we are interested in comparing analytical and numerical solutions in the regime
of total reflection for a cold electron plasma ($T_e\ll m_ec^2$) under the assumption of immobile
ions.  In this regime a standing wave is formed by the interference of
the incoming and reflected pulses, while penetration of the pulse into
the plasma is limited to the skin depth. Using cold fluid theory,
analytical solutions can be derived that describe the quasistationary
state reached by the plasma and electromagnetic
fields~\cite{marburger1975,goloviznin2000,cattani2000}.  Here we
follow the notation of Ref.~\cite{Siminos}.

In our simulations, we consider a laser pulse incident from vacuum $(x<0)$ on a
semi-infinite plasma slab $(x>0$). The initial distribution function
of the electrons is
\begin{equation}
   f(x,p_x)=
   \frac{n_0}{\sqrt{2\pi T_em_e}}\exp\left(-p_x^2/2m_eT_e\right) \Theta(x)
\end{equation}
where $\Theta(x)$ is the Heaviside function.
We introduce the normalized vector potential of the incoming laser pulse to be
$\g{a}_\mathrm{L}(x,t)=|q_e|\g A(x,t)/m_ec$. For circularly polarized light
\begin{align}
   \MoveEqLeft \g a_L(x,t)= \nonumber\\
   &a(t-x/c)\left[\cos\left(\omega t-kx\right)\hat y+\sin\left(\omega t-kx\right)\hat z\right],
\end{align}
where $\hat y$ and $\hat z$ denote unit vectors forming an orthonormal basis in the
plane transverse to the laser propagation direction. The envelope is
\begin{equation}
a(t-x/c)=
\begin{dcases}
(a_0/\sqrt{2})\sin^2\left(\frac{\pi}{8}(t-x/c)\right) & \text{if }
t<4T  \\
a_0/\sqrt{2} & \text{otherwise},
\end{dcases}
\end{equation}
where $T$ is the laser period and $a_0$ is the incident laser field amplitude. The laser pulse
is imposed as a boundary condition for the transverse magnetic field
on the left side of the simulation box, which is obtained by taking
the derivatives $\g B=\nabla \times \g A$.

The analytical theory is based on the assumption that the plasma-vacuum interface is pushed up to a point $x_b$ where the
ponderomotive force $-m_ec^2\partial \gamma/\partial x$ of the laser
pulse is balanced by the electrostatic field $q_eE_x$ due to charge
separation. Assuming that an equilibrium has been reached, $p_x=0$ and the relation between
the Lorentz factor and the normalized laser amplitude is $\gamma(x)=\sqrt{1 +
  a^2(x)}$. In the following, we express length in the inverse wave number $k^{-1}=c/\omega$ and density in terms of the critical density.
Letting  $a_b$ denote the value  of the vector potential
(envelope) at the equilibrium point $x_b$, it can be shown that
\begin{equation}
\frac{2a_0^2+a_b^4}{1+a_b^2}=2n_0\left(\sqrt{1+a_b^2}-1\right).
\end{equation}
This relation can be solved numerically for $a_b$ and the value $x_b$
is then calculated from
\begin{equation}
x_b=\frac{a_b}{n_0}\sqrt{\frac{2a_0^2-a_b^2}{1+a_b^2}}.
\end{equation}
Using these values for $x_b$ and $a_b$, the normalized vector
potential in the vacuum and plasma regions become
\begin{equation}
a(x)=\sqrt{2}a_0\sin\left[\arcsin\left(\frac{a_b}{\sqrt{2}a_0}\right)-(x-x_b)\right]
\end{equation}
and
\begin{equation}
a(x)=\frac{2\sqrt{n_0(n_0-1)}\cosh[(x-x_0)/\lambda_s]}{n_0\cosh[(x-x_0)/\lambda_s]-(n_0-1)},
\end{equation}
respectively, where $\lambda_s= 1/\sqrt{n_0-1}$ is the skin depth and
$x_0$ is determined by ensuring the continuity of the vector potential
at $x=x_b$. The electron density is finally calculated from
$n_e=n_0+\partial^2\gamma/\partial x^2$ \cite{Siminos}.

Figure~\ref{fig:quasistationary} shows, that the analytical
solutions for the electron density and vector potential given above
agree well with solutions obtained using {\sc veritas}, for a
plasma of density $n_0=2n_c$ and the four different laser intensities
$a_0=0.25$, 0.50, 0.75 and 1.00. In the numerical simulations, the
semi-infinite cold plasma was represented by a finite plasma slab of
length $5\lambda$. The dimensions of the simulation box were
$[-3,7]\lambda\times[-8,8]m_ec$. The mesh consisted
of five levels with refinement ratio $r=2$ and, on the coarsest
  level, $n_x=76$, $n_p=50$ points in the spatial and momentum
direction respectively. The temperature of the electrons was
taken to be $T_e=5\cdot 10^{-4}m_ec^2$. The finite temperature and
additional heating due to the finite rise time of the laser pulse
results in a slightly less peaked structure in the electron density at
the vacuum plasma interface compared to analytical theory which
neglects these effects.  However, reducing the temperature
significantly would increase the number of points in momentum space that
are needed to resolve the $p_x$ dependence of the distribution
function.

\subsection{Comparison to PIC simulations in different interaction regimes\label{sec:benchmark2}}

We performed simulations in order to study the performance of our code 
in three different interaction regimes: underdense plasma,
the relativistic self-induced transparency (SIT) regime and the hole-boring regime.
For relativistic laser pulses with $a_0\gtrsim1$ propagating in infinite plasma the classical
critical density $n_c=\omega^2m_e\epsilon_0/e^2$, introduced in
Sect.~\ref{sec:benchmark1} has to be modified in order to take into
account the dependence of the electron effective mass on the
$\gamma$-factor.  Using the normalizations of
Sect.~\ref{sec:benchmark1} for CP pulses and taking into account
conservation of canonical momentum, this leads to the relativistic
critical density $n_c^{\rm eff}=\sqrt{1+a_0^2/2}\, n_c$~\cite{akhiezer1956, kaw1970}.  The possibility
for propagation of a relativistic pulse in a classically overdense
plasma is known as self-induced transparency. However, as we
have seen in Sect.~\ref{sec:benchmark1}, for semi-infinite plasma -- a local  density peak is
formed at the plasma-vacuum interface, leading to a critical density departing from
$n_c^{\rm eff}$~\cite{goloviznin2000,cattani2000}.  The situation is
complicated by kinetic effects~\cite{Siminos} and ion
motion~\cite{Weng,Siminos2} and the threshold for transition from the
transparent to the opaque regime has to be determined numerically.  In
particular, when ion motion effects are taken into account one studies
the transition between SIT and the so-called hole-boring
regime~\cite{macchi2005,klimo_PRSTAB_2008,robinson_NJP_2008,yan_PRL_2008,naumova_PRL_2009}
in which the ions are accelerated in the charge-separation-induced
electrostatic field and the whole plasma-vacuum interface recedes
deeper into the plasma.

\begin{figure}[!tb]
    \centering
    \includegraphics[width=0.92\columnwidth]{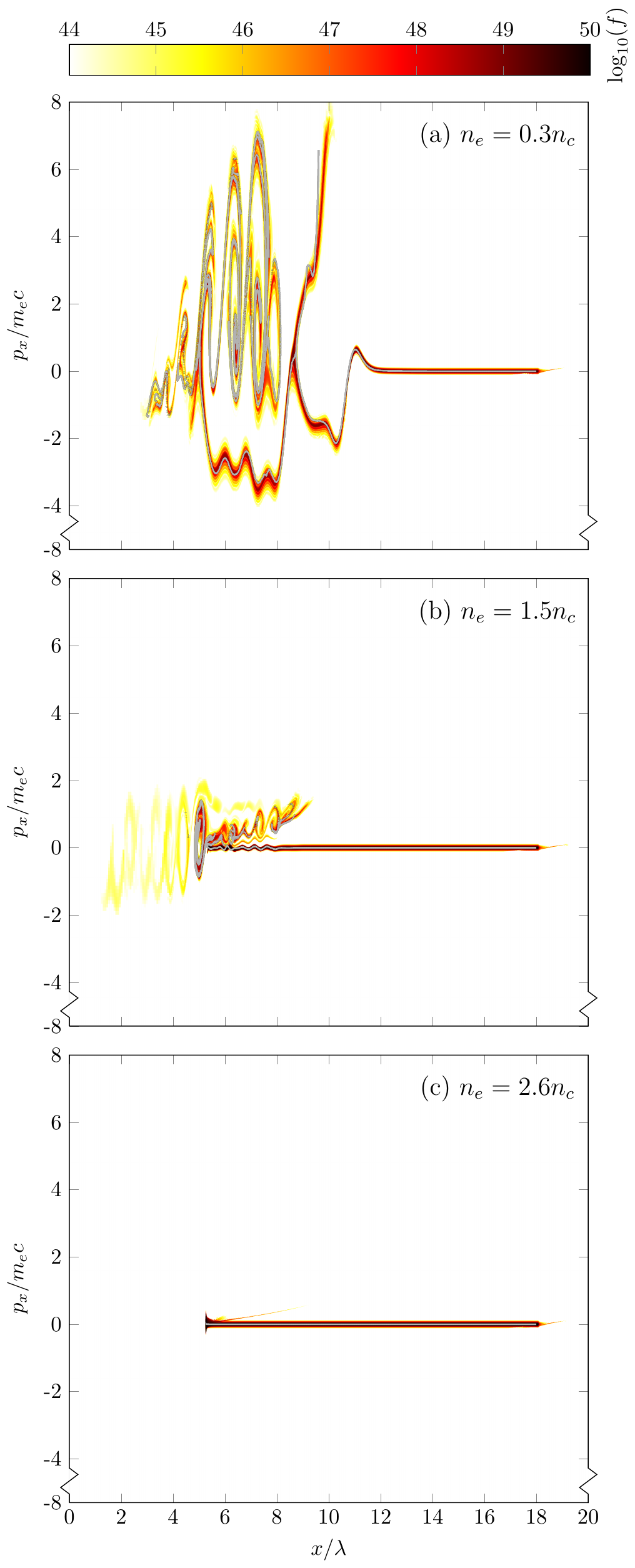}
    \caption{\label{fig:phase_space}The distribution function for electrons, calculated by \textsc{veritas}, at time $t=15T$ after      the front of the laser pulse has reached the plasma, for the densities (a) $n_e=0.3n_c$ (in the underdense plasma regime), (b) $n_e=1.5n_c$ (in the SIT regime) and (c) $n_e=2.6n_c$ (in the hole-boring regime). The corresponding distribution function (grey) from PIC simulations is also shown.}
\end{figure}

We performed simulations in order to study the interaction of plasmas with different densities and a laser pulse with  normalized laser intensity $a_0=2$. Figure~\ref{fig:phase_space} shows the distribution function for
electrons at time $t=15T$ after  the front of the laser pulse has
reached the plasma, for the densities $n_e=0.3n_c$ (in the underdense plasma regime),
$n_e=1.5n_c$ (in the SIT regime) and
$n_e=2.6n_c$ (in the hole-boring regime). The $n_e=0.3n_c$ case is in the regime where the relativistic Raman and modulational instabilities merge~\cite{Guerin1995,Guerin1996}
as evidenced by particle trapping and acceleration, the $n_e=1.5n_c$ case develops electron vortices on the front side which is in agreement with dynamics of the SIT regime \cite{Ghizzo2007,Siminos,Siminos2} and the $n_e=2.6n_c$ is identified with characteristics of the hole-boring regime. The Figure also shows the distribution function calculated by the PIC code (grey) at the same instant of time. We observe remarkable agreement between \textsc{epoch} and \textsc{veritas} simulations. 

In the \textsc{veritas} simulations presented above, the dimensions of the simulation
box were $[0,20]\lambda\times[-8,8]m_ec$ for electrons
and $[0,20]\lambda\times[-400,400]m_ec$ for
ions. Furthermore, the mesh consisted of five levels with refinement
ratio $r=2$ and $n_x=292$, $n_p=150$ points in the spatial and momentum
directions respectively, on the coarsest level. The temperatures of both electrons and ions were taken to be $T_e=T_i=5\cdot
10^{-4}m_ec^2$. For \textsc{epoch}, a grid resolution of 200 cells per wavelength was used, with each cell spawning 8000 third order B-spline particles.

\subsection{Efficient modelling of laser plasma interaction\label{sec:performance}}

The performance of the adaptive solver  {\sc veritas} is evaluated
by comparing to the performance of using a uniform grid, i.e.\ when the adaptive solver is operated with a single
level.  To identify how the performance depends on the problem, we consider the three cases $n_e=0.3n_c$, $1.5n_c$ and $2.6n_c$, which were described in
subsection \ref{sec:benchmark2}. The case with lowest density has the most complex electron dynamics and will hence be the most computationally heavy problem per time step (having the most degrees of freedom) -- even with an adaptive mesh. The higher densities have simpler and more localized electron dynamics, which demands a fine mesh with high resolution in only a small region of phase space.

In each case, simulations were performed for ten laser periods once the front of the
laser pulse had reached the vacuum-plasma interface.  For the adaptive
solver, we used the adaption parameters
$w_{x,1}=w_{p_x,1}=0.5\cdot r^{i/2}/f_\text{max}$,
$w_{x,2}=w_{p_x,2}=0.5\cdot r^{2i}/f_\text{max}$,
$w_f =6.25\cdot 10^{-6}\cdot r^i/f_{\text{max}}$ and
$\delta_{\rm thres}=10^{-8}$. These result in a relatively strongly refined mesh,
even in regions with small values of the distribution
function. Furthermore, the mesh was adapted every 20 time steps. In
all cases, the code was compiled using the Intel compiler with -O3
optimizations and ran on a desktop PC using an
Intel Xeon E3-1231 (3.4 GHz, 4 cores) CPU.

The performance of the adaptive solver can be quantified by the reduction of its memory footprint as well as the speedup compared to a uniform case. Figure \ref{performaceFig}(a) compares the memory footprints against system time, and Figure \ref{performaceFig}(b) shows the time it takes to perform a single time step, as a function of the time during the simulation (measured in laser periods).
The reduction in memory footprint is time varying, but is at least a factor of $3$ in all of the adaptive cases compared to the non-adaptive case.
In all cases we observe a speedup of at least $4$x (worst case, however for 50\% of the simulation time we see $\sim7$x). The difference in time per time step is caused by a larger number of degrees of freedom in the lower density cases, however this difference is modest.

\begin{figure}[bt]
    \includegraphics[width=\columnwidth]{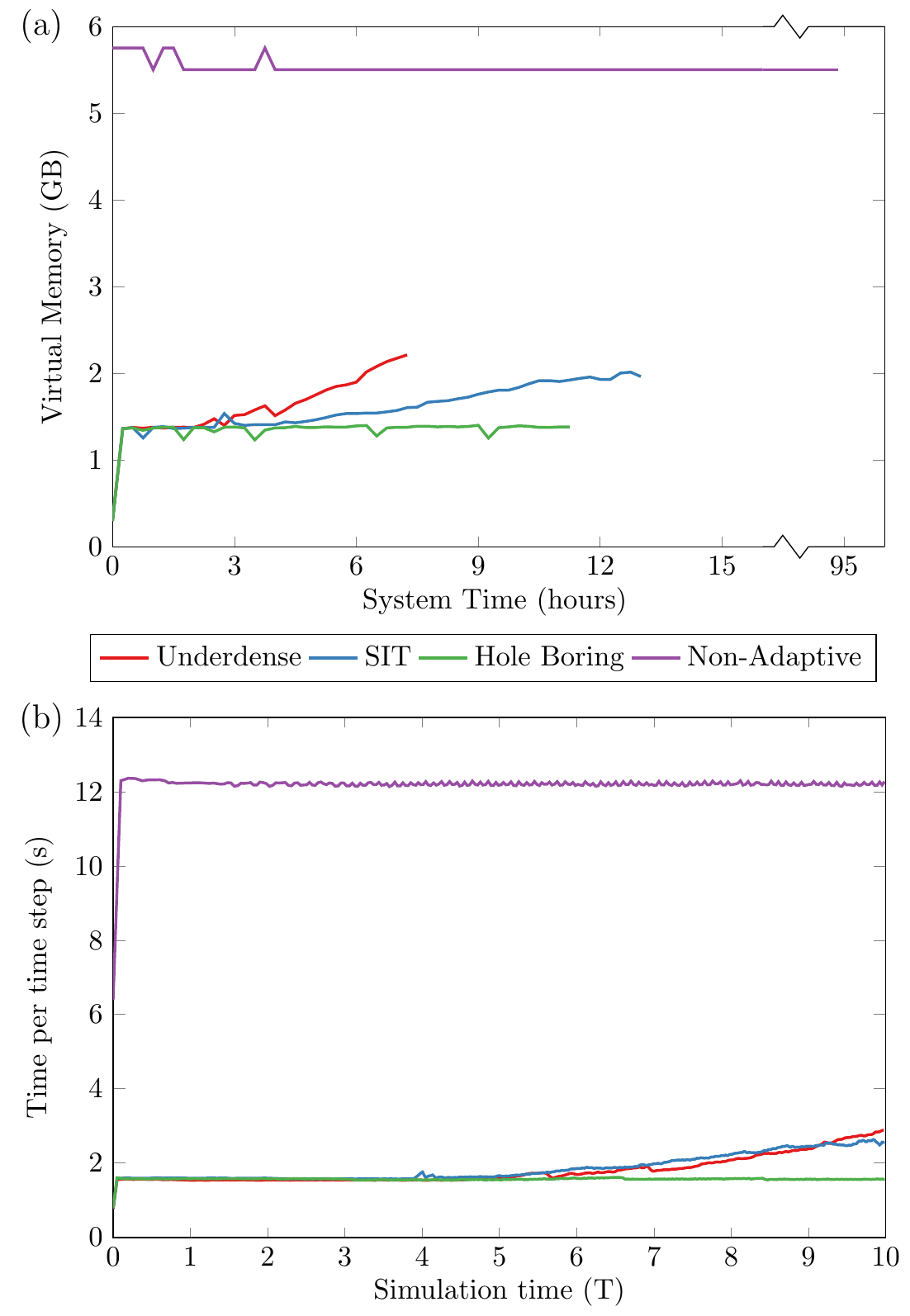}
    \caption{\label{performaceFig}(a) Memory footprint as a function of system time and (b) the time it takes to advance the solution a time step as a function of time in laser periods ($T$), when using an adaptive mesh to simulate a circularly polarized laser pulse impinging on plasmas with densities $n_e=0.3n_c$ (underdense plasma), $n_e=1.5n_c$ (SIT) and $n_e=2.6n_c$ (Hole Boring), as well as when using a uniform (Non-Adaptive) mesh.}
\end{figure}

To compliment the results in Figure \ref{performaceFig}, we  make a more in depth comparison of run time for the hole-boring case $n_e=2.6n_c$, with and without adaption. The total runtime and division of the computational work between
different parts of the solver is reported in Table
\ref{tab:runtime}. The runtime has been divided into four categories:
(1) a time step category which includes the time for advancing the
distribution function, (2) a category for the calculation of charge and
current densities, (3) a category for the regridding procedure and (4) a
category for the calculation of the electrostatic potential. The greatest proportion of the runtime, in all cases, was spent in one of these categories.

\begin{table}[tb]
\caption{\label{tab:runtime}A summary of the computational costs in
  key parts of \textsc{veritas}, with and without adaptive mesh
  refinement. Here, we limit the meaning of \textit{time step} to only
  include the contribution from time advancement of the distribution
  functions and exclude any electromagnetic field or
  potential updates.}
\begin{tabular}{l r@{.}l r@{.}l r@{.}l r@{.}l  }
\hline\hline

Category & \multicolumn{4}{c}{Uniform [s, (\%)]} & \multicolumn{4}{c}{AMR [s, (\%)]} \\ \hline
Time step           & 10360&1 & (32&6)  & 1514&1 & (40&5) \\
Charge \& currents  & 21228&7  & (66&8)  & 2000&1 & (53&5) \\
Update potential    & 127&1   & (0&4)   & 149&5  & (4&0) \\
Regridding          & 0&0     & (0&0)   & 56&1   & (1&5)  \\ \hline
Total               & 31779&4 & (100&0) & 3738&5 & (100&0)\\ \hline\hline
\end{tabular}
\end{table}

In both the uniform and adaptive cases, most of the run time is spent either on time stepping or calculating the charge and current.
The contribution to the run time from the solver for the transverse electromagnetic fields is marginal compared to the time spent in the above regions. Furthermore, we notice that the work load distribution is mostly unaffected by the use of adaption.

Notice that the computational cost associated with regridding was modest ($<2\,\%$). However,  there is a trade-off between choosing small adaption parameters,
resulting in a large number of points in the mesh, and high frequency
for adaption, resulting in large computational costs for regridding.
For example, in Ref.~\cite{Hittinger2013} a higher rate for adaption
is used, with the consequence that a larger amount of the
runtime is spent in regridding procedures. Optimal choices for the
adaption parameters and regridding rate is an area to investigate to
further optimize the accuracy and speed of the adaptive code.

\section{Conclusions}
The study of the Vlasov-Maxwell system of equations has long been of
interest for modelling collisionless plasma dynamics. Eulerian methods
are suitable for cases where high resolution of the distribution
function is needed, but have so far been restricted by the
computational expense. Here, an open source adaptive solver for the
relativistic Vlasov-Maxwell system in one dimension, \textsc{veritas},
is presented. We have successfully demonstrated the
    solvers capabilities, leading to a speed-up of a factor
  of $7$x in a typical scenario.  It is shown that the adaptive approach
  is well suited for problems where a small component of the electron or ion populations
  are accelerated to high velocity.

The discretization of the Vlasov-equation in \textsc{veritas} is based
on a high-order finite volume method, implementing the flux corrected
transport algorithm to limit spurious oscillations in the
  distribution functions, in the presence of steep gradients. The
use of efficient limiters is of critical importance to obtain a stable
numerical scheme over a wide parameter range for self-consistent
simulations with ultra intense fields, such as laser-plasma
interactions.

The reduction of runtime and memory footprint using adaption comes from the reduction in the total number 
of cells in which the computation is performed, although with some additional
overheads related to more complicated mesh structure and calculation of charge and
current densities.
In the cases and parameter sets considered in this paper, the overheads due to adaption were found to be minor. 
However, performance of the adaptive solver is
also connected to the optimization of adaption parameters, e.g., the
efficiency of the covering of marked cells using rectangles, and the
threshold for as well as form of the refinement indicator. The optimal
choices of these parameters are problem dependent.


Note that the local character of the explicit discretization
in \textsc{veritas} makes it well suited for parallelization,
resulting in an efficient tool for simulations of moderate laser
intensities (with $a_0\sim 1$).  However, in simulations of
interaction of a plasma with ultra-relativistic fields (with $a_0\gg
1$), performance for explicit methods is strongly limited by
CFL-restrictions on the time step.  Extensions to higher field
strengths may therefore benefit from a more robust approach
such as an explicit local time-stepping method \cite{Grote2010} or the use of semi-Lagrangian methods which put fewer restrictions on the timestep \cite{Shoucri2013}.
A third option may be an implicit treatment of the momentum derivatives in the Vlasov-equation, which would remove the severe scaling of the CFL-restriction due to the field
strength; although this method will pose challenges for a parallel implementation. 
 The small number of degrees of freedom in the adaptive mesh could be a significant advantage for the implicit approach.

 Finally, concerning the extention of \textsc{veritas} from a 1D1P code to 2D2P: all numerical methods presented here are equally applicable. Based on the performance of the 1D1P code, we are optimistic about the potential performance gains that can be achieved by the use of an adaptive mesh in 2D2P. On the other hand, some further investigation is still needed with regard to the scaling of overheads and effect of managing the more complicated mesh structure in a distributed memory context.

\section*{Acknowledgements}
The authors are grateful to S Buller, O Embr\'eus, E Highcock, S
  Newton, J Omotani, I Pusztai and A Stahl for fruitful discussions.
This work was supported by the Knut and Alice Wallenberg Foundation
and the European Research Council (ERC-2014-CoG grant 647121).  The
simulations were performed on resources at Chalmers Centre for
Computational Science and Engineering (C3SE) provided by the Swedish
National Infrastructure for Computing (SNIC).
EPOCH was developed under UK EPSRC grants EP/G054950/1, EP/G056803/1, EP/G055165/1 and EP/M022463/1.

\section*{Author contribution statement}

BSW and TCD developed the code and wrote the majority of the article. ES contributed with expertise on cold-fluid models and PIC-simulations. TF is the group leader of the Electromagnetic Field Theory group at Chalmers, hosting BSW, TCD and ES.

\bibliographystyle{epj}
\bibliography{SvedungWettervik}

\end{document}